# Structural, dielectric, ferroelectric and electrical properties of lead-free $Ba_{0.9}Sr_{0.1}Ti_{0.9}Sn_{0.1}O_3$ ceramic prepared by sol-gel method.

S. Khardazi[a*], H. Zaitouni[a], B. Asbani[b], D. Mezzane[a,b], M. Amjoud[a], E. Choukri[a], S. Terenchuk[c], A. Erramli[a], Y. Gagou[b]

[a] *IMED-Lab, Cadi Ayyad University, Marrakesh, 40000, Morocco*
[b] *LPMC, University of Picardy Jules Verne, Amiens, 80039, France*
[c] *Kiev National University of Construction and Architecture, Faculty of Automation and Information Technologies, Department of Information Design Technologies and Applied Mathematics, Vozdukhoflotsky Avenue, 31, 03037, Kiev, Ukraine*

**Abstract**

Lead-free $Ba_{0.9}Sr_{0.1}Ti_{0.9}Sn_{0.1}O_3$ (BSTSn) ceramic was prepared by the sol-gel method. X-ray diffraction measurement at room temperature reveals the single-phase formation with tetragonal structure refined in the P4mm space group. Raman spectroscopy analysis at room temperature was used to confirm the structure of BSTSn ceramic sintered at 1420°C during 5h. The dielectric measurements were examined in the temperature range [-90 – 250°C] and the frequency range of [100Hz – 1MHz]. The maximum dielectric constant was found to be ~5484 (1kHz) at room temperature. The diffuse phase transition characteristic at 1kHz was ascribed by the modified Curie-Weiss law. Macroscopic polarization – field (P – E) hysteresis loops recorded at room temperature (RT) confirm the ferroelectric nature of BSTSn ceramic. Impedance spectroscopy analysis of BSTSn sample in the temperature range 300 – 380°C reveals the presence of two relaxation contributions originated from grains and grain boundaries. Activation energies of grain and grain boundaries resistances were deduced in order to determine the conduction mechanism of the studied sample.

*Keywords:* dielectric; lead-free ceramic; sol-gel; ferroelectric; impedance spectroscopy.

* Corresponding author. Tel.: +212608219671.
  *E-mail address:* khardazzisaid@gmail.com

## 1. Introduction

During the past few decades, ferroelectric materials with perovskite structure have attracted much attention in several studies [1,2]. Especially, those exhibit large permittivity and piezoelectricity are extensively used in numerous devices, such as capacitors, piezoelectric sensors, and actuators [3]. Lead zirconate-titanate [Pb(Zr,Ti)O$_3$ (PZT)] present excellent dielectric, ferroelectric and piezoelectric properties [4]. This property makes them very suitable for various applications in electronic devices such as piezoelectric actuators, capacitors, sensors, etc [5]. This kind of material is limited due to the environmental impact induced by the toxicity nature of lead [4]. Barium titanate (BT) is an interesting alternative to lead-based materials due to its physical properties for developing environment-friendly materials [6, 7]. Numerous studies have been conducted to modify the BT system to enhance its dielectric and ferroelectric properties, versality and to extend the domain of its application and functionality, including the electrocaloric properties and the energy storage capacities [8–12]. Among this family, we found Barium Strontium Titanate Ba$_{1-x}$Sr$_x$TiO$_3$ (BST), which is formed by the substitution of Ba$^{2+}$ by Sr$^{2+}$ in the A-site of BT lattice. BST system has drawn ever increasing interest, because of its applications in dynamic random-access memory, tunable microwave devices, piezoelectric transducers and multilayer ceramic capacitors [1, 13–18]. Furthermore, it has a high dielectric constant, greater polarization, and a high dielectric breakdown strength, making BST ideal for high energy storage capacities. On the other hand, we found Barium Stannate Titanate (BTS) designed by the substitution of Ti$^{4+}$ by Sn$^{4+}$ in the BT perovskite structure. Because the curie temperature of BTS system could be widely shifted by changing the wt% of tin content, this solid solution can be used in various applications such as microwave phase shifter, actuator and capacitor [6, 19–21]. At present, few works have investigated the effect of simultaneous substitution of Sr$^{2+}$ and Sn$^{4+}$ in the A and B-sites of BT lattice, respectively [22, 23]. To synthesize this type of perovskite materials, various methods were employed such as solid-state technique, semi-wet method and sol-gel method [23–27]. Compared to conventional methods, the sol-gel route offers many advantages including simplicity, high control of stoichiometric composition of the phase composition, lower processing temperature and produce fine particles sizes of the powder. Moreover, the powders derived from sol-gel process are advantageous to the homogenous distribution of the composition and the microstructure, dense and better characteristics of ceramics [28, 29].

In this research work, Ba$_{0.9}$Sr$_{0.1}$Ti$_{0.9}$Sn$_{0.1}$O$_3$ bulk ceramics (abbreviated BSTSn) were prepared for the first time using the sol-gel method and their structure, microstructure, dielectric, ferroelectric and electric properties were investigated.

## 2. Experimental procedure

The Ba$_{0.9}$Sr$_{0.1}$Ti$_{0.9}$Sn$_{0.1}$O$_3$ (BSTSn) ceramic was prepared by the sol-gel method as described in the chart below (Fig. 1). High purity of barium acetate Ba(CH$_3$COO)$_2$, strontium acetate Sr(CH$_3$COO)$_2$·1/2H$_2$O, tin chloride dehydrate SnCl$_2$·2H$_2$O, and Titanium Isopropoxide (C$_{12}$H$_{28}$O$_4$Ti) were weighed according to stoichiometric ratios (Table 1). Acetic acid CH$_3$CH$_2$OOH and 2-methoxyethanol (C$_3$H$_8$O$_2$) were used as solvent agents. Stoichiometric quantities of barium acetate and strontium acetate were mixed and dissolved in acetic acid. Appropriate amounts of tin chloride dehydrate and Titanium Isopropoxide were mixed with 2-methoxythanol under stirring for 1h. The pH~7 of the solution was adjusted by adding a Required quantity of Ammonia leading to a transparent sol. The obtained solution was heated to 80°C for 1h in the hot air oven, leading to white gel. The dried powder was calcined for 5h at 1000°C and pressed into disks using the uniaxial hydraulic press. The resulting pellets were then sintered at an optimized temperature of 1420 °C for 5h. It is interesting to note that the sintering conditions used in the present study were taken based on the density measurements using Archimedes method.

The X-ray diffraction (XRD) pattern was registered at room temperature using the Panalytical X-Pert Pro under Cu-Kα radiation with λ ~ 1.540598 Å. The grain morphology of the sintered ceramic was observed by using the TESCAN VEGA3 Scanning Electron Microscope (SEM). The dielectric measurements were measured in the frequency range 100 Hz – 1 MHz and temperature interval from -90 °C to 250 °C by the Solartron Impedance Analyzer SI-1260. Room temperature polarization versus electric field (P-E) hysteresis loops were performed using a ferroelectric test system ( PolyK Technologies State College, PA, USA) at 200Hz.

Table 1. Masses of raw materials used to prepare $Ba_{0.9}Sr_{0.1}Ti_{0.9}Sn_{0.1}O_3$ ceramics.

| Mass of raw material/g | | | |
| --- | --- | --- | --- |
| $Ba(CH_3COO)_2$ | $Sr(CH_3COO)_2 \cdot 1/2H_2O$ | $SnCl_2 \cdot 2H_2O$ | $C_{12}H_{28}O_4Ti$ |
| 5.1763 | 0.4632 | 0.5081 | 5.7600 |

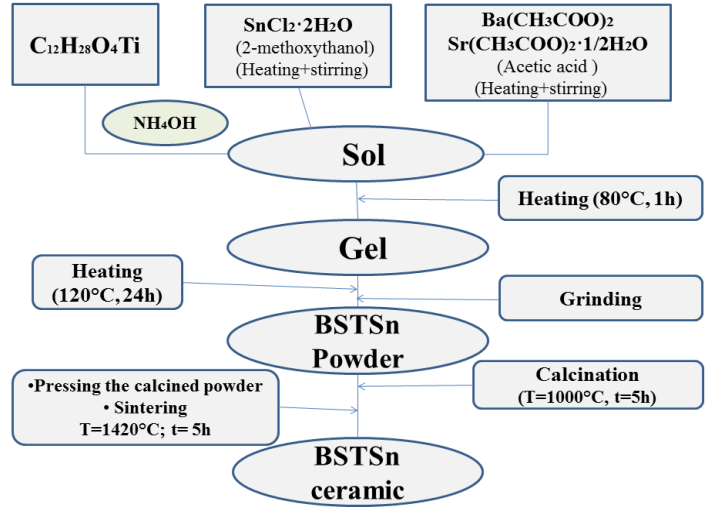

Fig. 1. Flow chart of the preparation of BSTSn ceramics by sol-gel method.

## 3. Results and discussion

### 3.1. X-ray diffractions

Room temperature X-ray diffraction patterns of BSTSn powder calcined at 1000°C for 5h is shown in Fig. 2. The sample shows a pure perovskite phase without any traces of impurities. The crystal structure of BSTSn ceramic was successfully refined in the tetragonal symmetry of the P4mm space group (by using full-prof software), suggesting the successful diffuseness of Sr and Sn into BT lattice to form a solid solution. By analyzing the shape of XRD lines, we have determined the average crystallite size of BSTSn powder using Debye Scherrer's formula [30]:

$$D = \frac{k\lambda}{\beta \cos\theta}, \quad (1)$$

where $k$ is the Scherrer constant (generally taken as 0.9), $\lambda$ is the wavelength of Cukα radiation, $\beta$ is the full width at half maximum (FWHM) of the broadened diffraction peaks and $\theta$ is the measured angle. From the calculations, the average crystallites size of BSTSn powder was found to be around 45 nm. Other structural parameters such as lattice parameters and atomic positions are listed in Table 2.

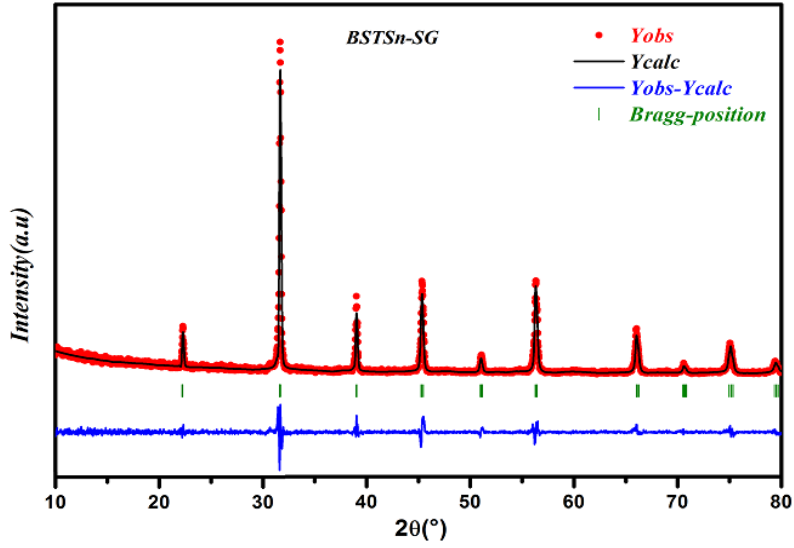

Fig. 2. X-ray diffraction patterns of $Ba_{0.9}Sr_{0.1}Ti_{0.9}Sn_{0.1}O_3$ powder calcined at 1000°C/5h.

Table 2. Structural details of BSTSn sample gathered from Rietveld refinement.

| Lattice parameter(Å) | Angle(°) | Space group | Atomic position(x, y, z) | | | | $\chi^2$ |
|---|---|---|---|---|---|---|---|
| a = 4.0029 | $\alpha = \beta = \gamma = 90$ | P4mm | Ba/Sr | 0.00000 | 0.00000 | -0.03855 | 1.34 |
| b = 4.0029 | | | Ti/Sn | 0.50000 | 0.50000 | 0.44897 | |
| c = 4.0110 | | | $O_1$ | 0.50000 | 0.50000 | -0.05816 | |
| | | | $O_2$ | 0.00000 | 0.50000 | 0.48410 | |

*3.2. Raman spectroscopy*

Fig.3 shows Raman spectra of BSTSn ceramic recorded at room temperature. From Fig. 3, two transverse optical modes [$A_1$ (TO)] and a longitudinal optical mode [$A_1$ (LO)] were observed in the range of 100-300 $cm^{-1}$. The detected bound at around 303 $cm^{-1}$ is a distinctive particular of the tetragonal phase and it is attributed to a combined mode [B1, E (TO + LO)]. This Raman bound corresponds to the asymmetrical vibration mode of $TiO_6$ octahedral. The frequencies of bout 516 and 724 $cm^{-1}$ are assigned to two mixed modes, [$A_1$, E(TO)] and [$A_1$, E(LO)] produced by Ba-O bands [31]. These results indicate the presence of tetragonal distortion in BSTSn ceramic.

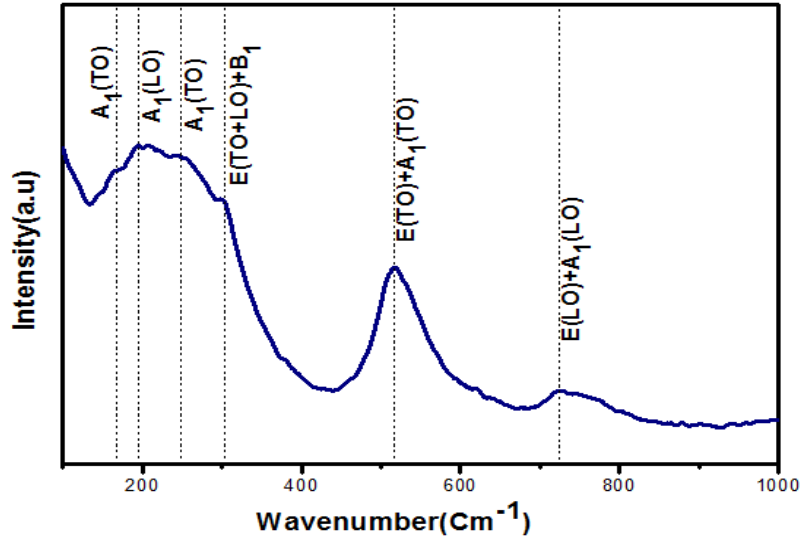

Fig. 3. Room temperature Raman spectra of BSTSn ceramic.

*3.3. Microstructure analysis*

Fig. 4 presents the SEM micrograph and grains morphology of BSTSn ceramic sintered at 1420°C/ 5h. The sample shows a dense microstructure with non-uniform grain size. The average grain size was determined using the Gaussian distribution of the grain and found to be around 6.80 ± 0.17 μm. Indeed, with sintering at such conditions, the BSTSn ceramic shows a pure, dense microstructure and small average grain size, which confirm the good quality of the ceramic for excellent electrical properties [32]. This result could be attributed to the fine crystallites obtained from the powder synthesized using sol-gel technic, as discussed in the X-ray diffraction section above. The fine particles obtained tend to agglomerate, resulting in a non-uniform and well-connected grain [33].

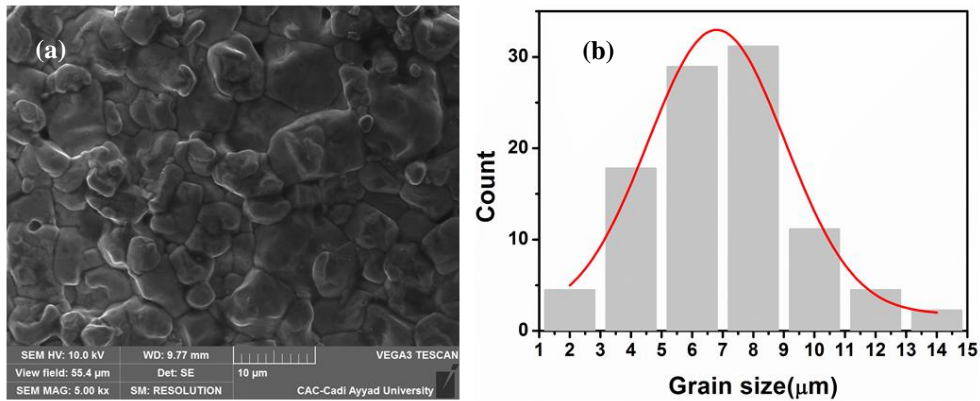

Fig. 4. (a) SEM micrograph and (b) grain size distribution of BSTSn ceramic sintered at 1420°C/5h.

*3.4. Dielectric properties*

Fig. 5 (a) and (b) show the temperature dependence of dielectric constant ($\varepsilon_r$) and dielectric loss (tgδ) of BSTSn ceramic in the frequency range 100 Hz – 1 MHz. The $\varepsilon_r$ increases with increasing temperature and reveals a maximum value ($\varepsilon_m$) of 5484 (1 kHz of frequency) at $T_C$ = 25 °C, which corresponds to the temperature phase transition from ferroelectric to paraelectric phase. These results are in good agreement with those reported in the literature for BSTSn ceramics [26]. The obtained $\varepsilon_m$ in this work is relatively low in comparison with those mentioned for BSTSn ceramics prepared by the conventionnel solid-state method, in which a high calcination temperature is adopted [23]. On the other hand, the relatively low value of $\varepsilon_m$ ~ 5500 could be attributed to the small grain size obtained for this composition by sol-gel method. One can see from Fig.5 (a) and (b), that both dielectric constant and dielectric loss are decreasing with increasing frequency. This decrease could be attributed to the decrease of the contributions of different polarizations at high frequencies [34].

In the paraelectric region, the dielectric response usually follows the Curie-Weiss law for (T > T$_C$):

$$\frac{1}{\varepsilon_r} = \frac{(T - T_0)}{C} \quad (2)$$

where $T_0$ and $C$ are the Curie-Weiss temperature and the Curie-Weiss constant, respectively. Fig.5 (c) display the temperature evolution of the inverse dielectric constant $\left(\frac{1}{\varepsilon_r}\right)$ at 1 kHz of BSTSn ceramic. Good agreement between experimental data and those calculated by using Eq. (2) is observed. The calculated parameters are listed in Table 3. The obtained value of $C \sim 10^5$ K confirms the displacive-type ferroelectric of BSTSn ceramic.

To explain the diffuser behavior of the phase transition, the parameter ΔTm is used to reveal the degree of diffuseness. This parameter was determined by using the following equation [35]:

$$\Delta T_m = T_{dev} - T_m \quad (3)$$

here, $T_{dev}$ is the temperature at which the dielectric constant begins to deviate from the Curie-Weiss law and T$_m$ is the temperature corresponding to the maximum dielectric constant.

The temperature dependence of the dielectric permittivity, above its maximum, has been extensively investigated as an important parameter to determine the character of phase transition in relaxor ferroelectrics. In this region, temperature-frequency dependence is not detected in $\varepsilon_r$ versus T curves for the majority of relaxors [36]. The diffuseness coefficient of a diffuse phase transition (DPT) has been extracted from the Eq. (4) below known as the modified Curie-Weiss law:

$$\frac{1}{\varepsilon_r} - \frac{1}{\varepsilon_m} = \frac{(T - T_m)^\gamma}{C^*} \quad (4)$$

where $\gamma$ is the constant that defines the transition diffuseness degree, $C^*$ is the modified Curie-Weiss constant and $T_m$ is the temperature corresponding to the $\varepsilon_m$. The characteristics of the phase transition depend on γ values. γ = 1 indicates the so-called ''conventional'' ferroelectric phase transition while when γ = 2, we have a ''complete'' DPT. Intermediate values of γ between 1 and 2 describe an "incomplete" DPT [37]. Fig. 5(d) presents the plot of $\ln\left(\frac{1}{\varepsilon_r} - \frac{1}{\varepsilon_m}\right)$ versus $\ln(T - T_m)$ at a selected frequency of 1 kHz. The adjustable parametres are resumed in Table 3. The obtained γ value of BSTSn sample was found to be 1.62 indicating an incomplete diffuse phase transition.

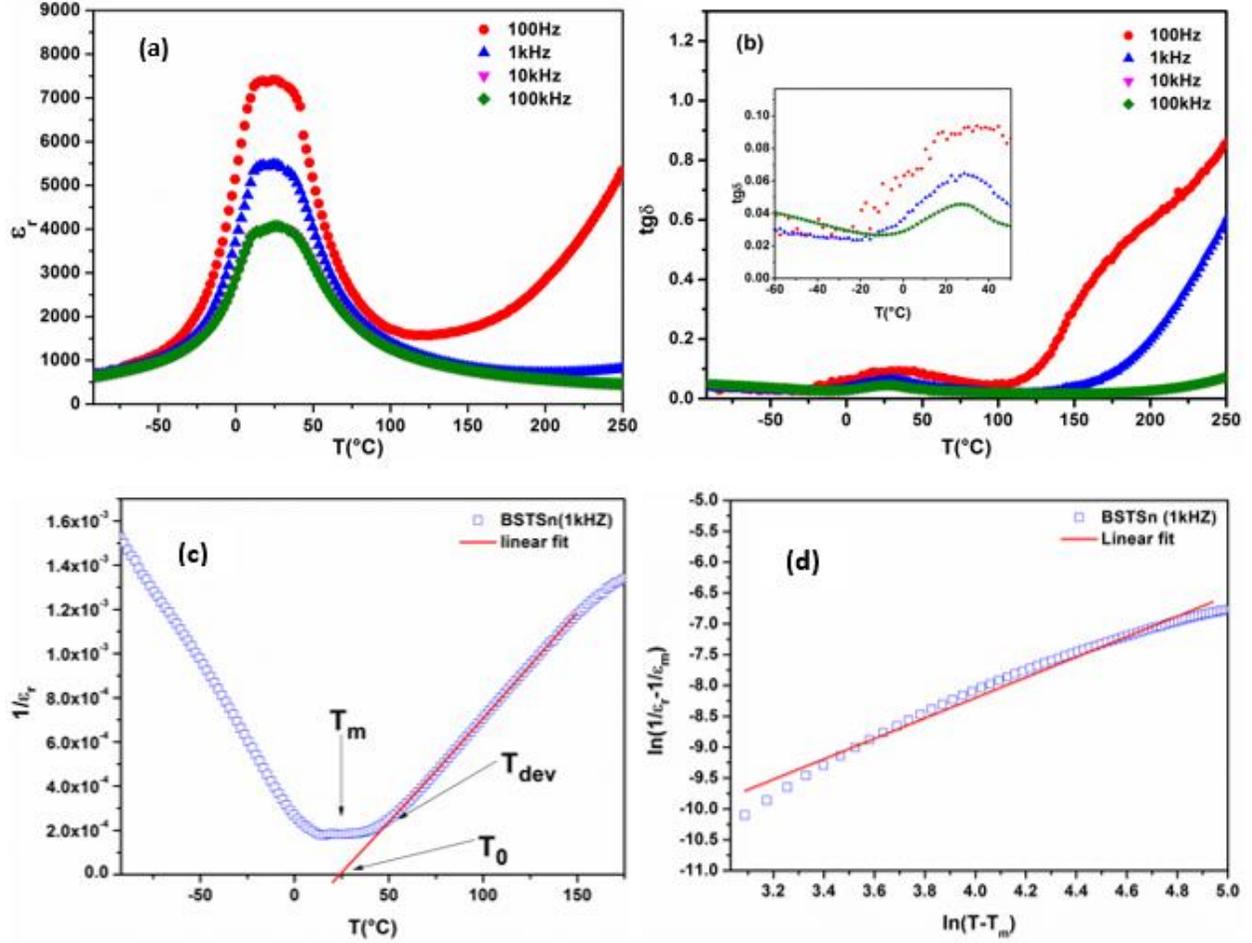

Fig.5 .(a) temperature- dependence of dielectric constant,(b) temperature- dependence of dielectric losses, (c) Thermal variation of $1/\varepsilon_r$, and (d) showing plot of modified Curie-Weiss law at 1kHz of BSTS ceramic sintered at 1420°C/5h.

Table. 3 Adjustable parameters obtained from Eqs. (2), (3) and (4) of BSTSn ceramic at 1kHz.

|  | $\varepsilon_m$ | $T_0$(°C) | $T_m$(°C) | $C \times 10^5$ | $T_{dev}$ | $\Delta T_m$ | $\gamma$ |
| --- | --- | --- | --- | --- | --- | --- | --- |
| BSTSn | 5484 | 26 | 23.6 | 1.04 | 52.81 | 29.21 | 1.62 |

## 3.5. Ferroelectric properties

In order to explore the ferroelectric properties of the BSTSn ceramic, the polarization-field (P – E) hysteresis loop was performed at RT and 200 Hz of frequency (Fig. 6). Typical ferroelectric P – E cycle is observed which confirms the ferroelectric nature of BSTSn ceramic. Indeed, the studied sample presents a very narrow hysteresis cycle with a suitable symmetric shape. It is found that BSTSn sample show high spontaneous ($P_s$) polarization value of 11.44 $\mu C/cm^2$ with low values of remanent polarization ($P_r$) and coercive field ($E_C$) of 2.94 $\mu C/cm^2$ and 0.53 KV/cm, respectively under an applied field of 23 KV/cm. These findings indicate that BSTSn ceramics are very promising for energy storage applications. It is worth to mentioning that the maximum applied field of BSTSn in our study is very high than the one applied for BSTSn ceramics prepared by the solid-state method [23]. This result demonstrates that the sol-gel method is suitable for synthesizing ceramics with moderate breakdown strength.

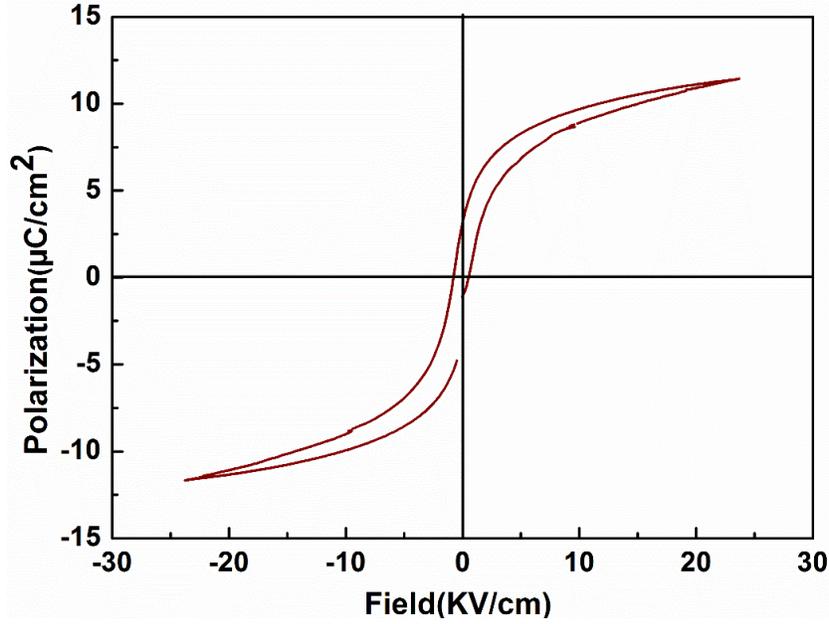

Fig. 6. Hysteresis loop of BSTSn ceramic obtained at room temperature.

*3.6. Complex impedance analysis*

In order to perform a thorough study of the electrical characteristics of BSTSn sample, the complex impedance spectroscopy (SIC) was used. Fig. 7 depicts the Nyquist plots (-Z″ versus Z′) of BSTSn sample at different temperatures. The impedance spectrum from RT up to 300 °C is not shown in the figure (Fig. 7) as they just demonstrated a straight line with a large slope suggesting the insulating behavior of the sample. Nevertheless, while increasing temperature (far away from $T_C$), the semi-circular arcs are formed. In the present case, our experimental data are located on two semicircles, that can be attributed to the grain and grain boundaries contributions at high and lower frequencies, respectively [38]. Moreover, the radius of both semicircles was found to decrease with increasing temperature indicating the increase in conductivity of the sample with temperature. The complex impedance spectra was well fitted by using Z-View software with an equivalent electric circuit consisting of two parallel resistance and a constant phase element (R//CPE) connected in series (red fitted line). The resistances of both grain ($R_G$) and grain boundaries ($R_{GB}$) at different temperatures were extracted from fitting results and plotted in Fig. 8 as a function of the inverse of temperature. The activation energy process ($E_a^R$) of both $R_G$ and $R_{GB}$ was investigated. The evolution of both resistances obey to the Arrhenius law given by:

$$R = R_0 \exp(E_a^R / k_B T) \qquad (5)$$

Where $R_0$ is the pre-exponential factor and $k_B$ is the Boltzmann constant. It was reported in the literature that the oxygen vacancies can be easily created during the sintering process at high temperatures in perovskite materials containing titanate [39]. For perovskite oxides, the activation energy for conduction in the ranges of 0.3–0.5 eV and 0.6–1.2 eV have been associated to the singly and doubly ionized oxygen vacancies ($V_O'$, $V_O''$), respectively [40]. One can see from Fig. 8 that the evolution of both $R_G$ and $R_{GB}$ in the studied range temperature [300 – 380°C] can be separated into two regions (T < 340°C) and (T > 340°C). Therefore, two activation energies were deduced for each contribution. The activation energies were found to be 0.45 eV and 0.87 eV for both grains and grain boundaries, respectively for (T < 340 °C). These values of activation energy suggest that the conduction process below 340 °C is governed only by the oxygen vacancies singly and doubly ionized. However, for T > 340°C, the activation energies were found to be higher than 2 eV for both contributions, which suggests that the conduction process could be assigned to the barium vacancies along with the oxygen vacancies [41].

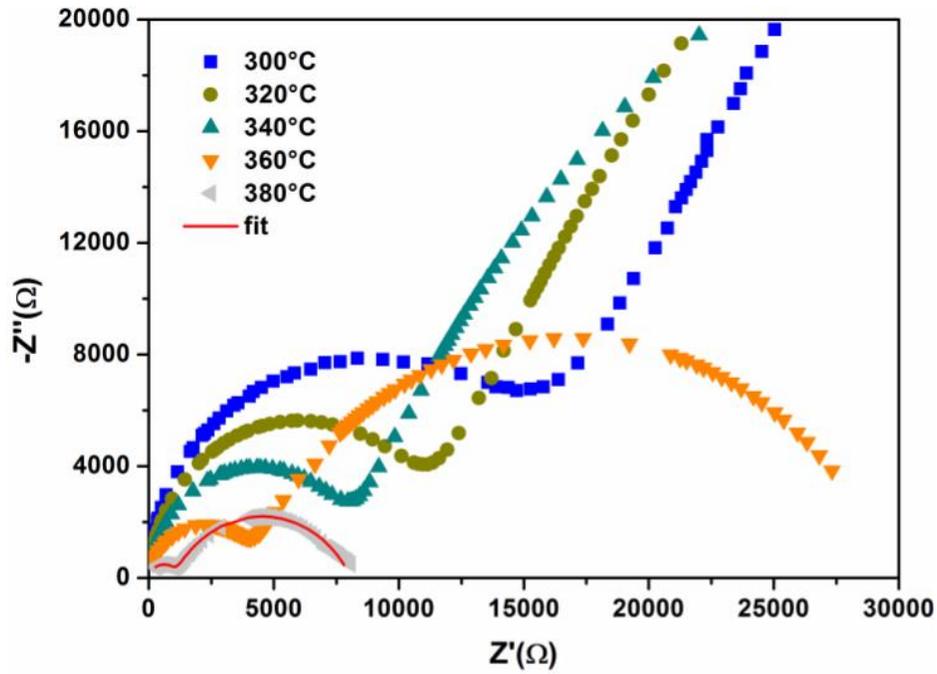

Fig. 7. Nyquist plot of BSTSn ceramic at various temperatures. The fitted curve is represented by a red line for representative temperature (380°C).

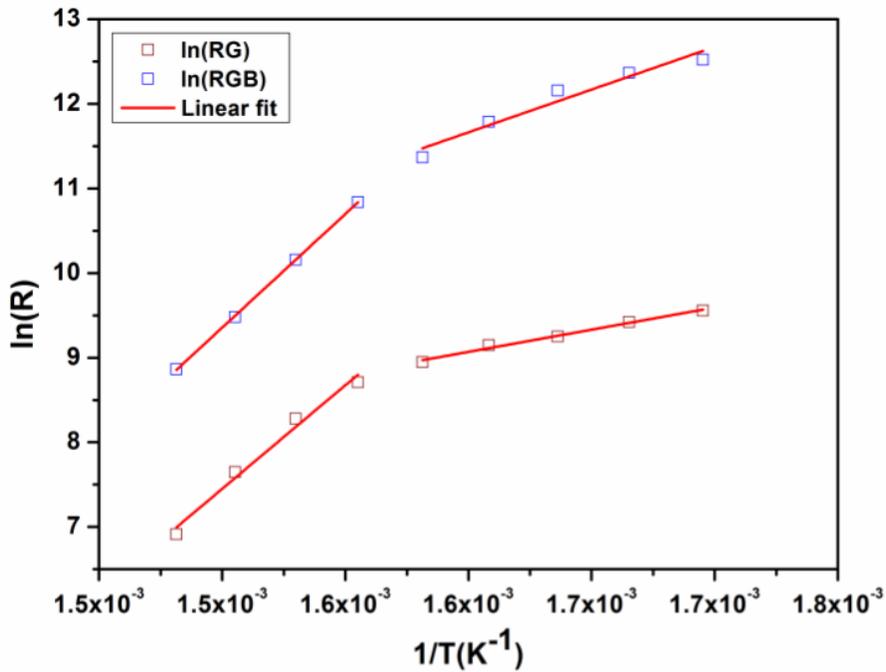

Fig. 8. Variation of the resistance of grain and grain boundaries as a function of the inverse of temperature for BSTSn ceramic.

## 4. Conclusion

In the present work, lead-free $Ba_{0.9}Sr_{0.1}Ti_{0.9}Sn_{0.1}O_3$ ceramic was successfully synthesized by the Sol-gel process. The XRD analysis showed a pure perovskite with a tetragonal structure refined in P4mm space group. The scanning electron microscopy showed a relatively large grain size ( 6.80 ± 0.17 μm) for BSTSn ceramic sintered at 1420°C. The dielectric analysis of BSTSn ceramic reveals a relatively high dielectric constant of ~5484 at room temperature with low dielectric losses less than 0.06. The modified Curie-Weiss law was used to study the diffuseness of BSTSn ceramic at 1kHz and suggests the incomplete diffuse nature of the transition. The ferroelectric nature of BSTSn ceramic was highlighted by P-E hysteresis loops and the values of the different parameters $P_s$, $P_r$ and $E_c$ were determined at RT. Different activation energies for conduction were observed in the temperature range 300 – 380°C. The obtained values reveal that the conduction mechanism in our BSTSn sample originated from the oxygen vacancies (singly and doubly ionized) below 340 °C and barium vacancies above 340 °C. Such properties make BSTSn ceramics a potential candidate for divers highly developed technological applications working at RT. For future research, the material selected in this study will be used as a target for synthesizing thin films by depletion laser technique.

**Acknowledgments**

The authors gratefully acknowledge the financial support of CNRST Priority Program PPR15/2015 and the European H2020-MSCA-RISE-2017-ENGIMA action.